\documentstyle[bezier,12pt]{article}
\def\grle{\raisebox{.25em}{$>$}\hspace{-.7em}\raisebox{-.25em}{$<$}\hspace
{.1em}}
\def\hil{{\cal H}}
\def\intreal{\int_{-\infty}^{\infty}}
\def\intrealp{\int_{0}^{\infty}}

\def\real{{\bf R}}
\def\rhs{{R.H.S.}}
\def\qm{quantum mechanics}
\def\Qm{Quantum mechanics}
\def\QM{Quantum Mechanics}
\def\qmal{quantum mechanical}
\def\zRm{{z_{R}}^{-}}
\def\zRsp{{z_{R}^{*}}^{+}}
\newtheorem{th}{Theorem}
\normalbaselines
\begin{document}
\hfill{\Large ASI-TPA/6/95}
\begin{center}
\vspace*{1.0cm}

{\LARGE{\bf  The Rigged Hilbert Space Formulation of Quantum Mechanics and
its Implications for Irreversibility }}

\vskip 1.5cm

{\large {\bf Christoph Schulte, Reidun Twarock }}

\vskip 0.5 cm

Institut f{\"u}r Theoretische Physik (A) \\
TU Clausthal \\
38678 Clausthal-Zellerfeld, Germany
\vskip 1.5cm

based on lectures by
\vskip 0.5cm

{\large {\bf Arno Bohm }}
\vskip 0.5 cm

CPP, Physics Department \\
University of Texas at Austin\\
Austin, TX 78764, USA
\end{center}
\vspace{1 cm}

\begin{abstract}
\Qm\ in the Rigged Hilbert Space formulation describes quasistationary
phenomena mathematically rigorously in terms of Gamow vectors.
We show that these vectors exhibit microphysical irreversibility, related to
an intrinsic \qmal\ arrow of time, which states that preparation of a state
has to precede the registration of an observable in this state.
Moreover, the Rigged Hilbert Space formalism allows the derivation of an
exact golden rule describing the transition of a pure Gamow state into a
mixture of interaction-free decay products.
\end{abstract}

\vspace{1cm}

\section*{Introduction}

The idea of using the Rigged Hilbert Space formulation to describe \qm\
already occurred in 1966 in \cite{a} and was later
elaborated in \cite{b} and \cite{c}.
This formalism recaptures all standard results of \qm\ and makes
the mathematical theory of Dirac's bras and kets rigorous.
Unexpectedly it was found \cite{c,d} that it also allows the description of
resonances and decaying states.
Resonances in scattering theory are usually defined as the poles of the
analytically continued $S$-matrix.
The Rigged Hilbert Space provides a means to describe these states by means
of vectors.
We call these vectors Gamow vectors.
Their time evolution is given by a semigroup instead of the usual unitary
one-parameter group \cite{d,e}.
This particular feature made people aware of a connection between Gamow
vectors and irreversibility.
For example Antoniou, Prigogine et.~al.\ who tried to obtain intrinsic
irreversibility on the quantum level \cite{f} advocated the use of this
method to explain irreversibility on the microphysical level \cite{g}.
It finally could be shown in \cite{h} that the occurrence of the two
semigroups is linked to a \qmal\ arrow of time which stems from the fact,
that states must be prepared before observables can be measured in them.

In section \ref{sec:RHS} we briefly describe the Rigged Hilbert Space
formalism of \qm.
A more detailed presentation is provided by the above references, e.~g.\
\cite{e}.

Section \ref{sec:BohrLudwig} is devoted to the derivation of a \qmal\ arrow
of time from the time shift between the preparation and registration
procedure.
This \qmal\ arrow of time is based on an earlier arrow of time formulated by
Ludwig and based on the Bohr-Ludwig interpretation of quantum mechanics
\cite{i} which however was only formulated for the experimental preparation
and registration apparatuses and not incorporated into quantum mechanics.
The reason for this was, that the standard quantum mechanics in Hilbert space
is reversible and cannot accommodate an arrow of time.

In section \ref{sec:mathform} a mathematical formulation of the \qmal\ arrow
of time for general scattering experiments is given.
We will cast it in a form which reappears in section \ref{sec:gamow}, where
the \qmal\ arrow of time in connection with Gamow vectors is discussed.
Finally, we consider decaying states and show how these results lead to an
exact golden rule.

\section{The Rigged Hilbert Space Formalism of \QM}
\label{sec:RHS}

By considering different topologies on a given linear space, we can produce
various complete topological spaces.
This idea is behind the Rigged Hilbert Space (\rhs) formalism, which
uses Gel'fand triplets of the form
\begin{equation}
\Phi \subset \hil \subset \Phi^{\times},
\end{equation}
where $\hil$ represents a Hilbert space with the well-known Hilbert space
topology $\tau_{\hil}$ and $\Phi$ a subspace with a nuclear topology
$\tau_{\Phi}$.
This topology is chosen to be stronger than the Hilbert space topology.

Furthermore, $\Phi^{\times}$ is the conjugate
space of $\Phi$, i.~e.\ the space of all $\tau_{\Phi}$-continuous antilinear
functionals $F(\phi )$, $\phi \in \Phi$, in the following denoted by
$F(\phi)=<\phi | F>$.
Notice, that $<\phi | F>$ with $F\in\Phi^{\times}$ is an extension of the
ordinary Hilbert space scalar product $(\phi, f)$ for $f\in\hil \cong
\hil^{\times}$.

The construction of the above triplet is such that $\Phi$ is $\tau_{\hil}$-
dense in $\hil$.
The space $\Phi$ will be used later to define the physical states.

In order to describe observables, we need the following triplet of linear
operators
\begin{equation}
\label{eq:conjop}
 A^{\dag}|_{\Phi} \subset A^{\dag} \subset A^{\times},
\end{equation}
where $A$ has to be a $\tau_{\Phi}$-continuous operator in $\Phi$ but is not
a closed operator in $\hil$.
The operator $A^{\dag}$ is the Hilbert space adjoint and $\bar{A}$ the
closure of $A$.
In general $A^{\dag}$and $\bar{A}$ are not $\tau_{\hil}$-continuous.
$A^{\times}$, defined by
\begin{equation}
<\phi | A^{\times} | F > = < A\phi | F >
\mbox{for all }\phi \in \Phi \mbox{ and } |F> \in \Phi^{\times}
\end{equation}
is a continuous operator in $\Phi^{\times}$, called the conjugate operator
of $A$.
Thus $A^{\times}$ is defined as the extension of $A^{\dag}$ to
$\Phi^{\times}$.
In this formalism, observables are represented by elements of an algebra of
$\tau_{\Phi}$-continuous operators whereas in the usual Hilbert space
formulation they are given by linear (unbounded) operators defined on $\hil$.
In the Hilbert space formulation, of \qm\ (pure) states are described by
elements of $\hil$; in the Rigged Hilbert Space formulation only elements of
$\Phi$ represent physically preparable states.

Generalized eigenvectors $F_{\omega}$ of the  operator $A$ with eigenvalue
$\omega$ are defined by
\begin{equation}
<A\phi | F_{\omega}>  \equiv < \phi | A^{\times} | F_{\omega} > =
\omega < \phi | F_{\omega}> \mbox{ for all } \phi\in\Phi,\,
F\in \Phi^{\times}.
\end{equation}
The Dirac kets are generalized eigenvectors in this sense and the expansion
of a physical state vector $\phi$ for an observable $H$ in terms of a basis
system of eigenkets, is given by the generalized basis vector expansion
which Dirac already used but which was proven only by the Nuclear Spectral
Theorem \cite{j}
\begin{equation}
\label{eq:nst}
\phi =
\int_{0}^{\infty} dE|E >< E | \phi> +
\sum_{E_{n}} |E_{n} )( E_{n}|\phi )
\mbox{ for every } \phi \in \Phi^{\times},
\end{equation}
where
\begin{equation}
H|E_{n})=E_{n}|E_{n}), \hspace{1ex}
H^{\times} | E > = E | E >
\end{equation}
with $E_{n}=-|E_{n}|$ being the discrete eigenvalues and $E$ elements of the
absolutely continuous spectrum of $H$.

The \rhs-formalism of \qm\ has several advantages in comparison with the
ordinary Hilbert space formalism:
\begin{enumerate}
\item The topology of the space $\Phi$ excludes states with infinite energy
which are elements of $\hil$.
\item Wave functions $\phi(E)$ --- omitting in this notation the dependence
on additional quantum numbers --- are always smooth functions.
In the Hilbert space formulation, the wave function is any function
$h(E)$ in the class of Lebesgue square integrable functions.
As $|\phi(E)|$ describes apparatus resolution, the Rigged Hilbert Space
formulation is closer to the experimental situation.
\item In contrast to the Hilbert space formulation, the algebra of
observables is in the \rhs\-formalism represented by continuous operators and
every essentially self-adjoint observable has a
complete set of generalized eigenvectors in $\Phi^{\times}$.
\item In addition to the Dirac kets, the \rhs\ contains generalized
eigenvectors with complex eigenvalues called Gamow vectors, which are
appropriate for the description of resonances and decaying states.
\item The Gamow vectors which are generalized eigenvectors of the
self-adjoint energy operator with complex eigenvalues have a  Breit-Wigner
energy distribution as necessary for the analysis of resonance scattering
experiments and are associated to the resonance poles of the $S$-matrix.
\item The time-evolution of the Gamow vectors is exactly exponential which
is the basis of microphysical irreversibility.
\end{enumerate}
It is shown in \cite[p.\ 48]{e}, that
\begin{equation}
L^{2}(\real) = \hil_{+}^{2} \oplus \hil_{-}^{2}
\end{equation}
where $\hil_{+}^{2}$ (respectively $\hil_{-}^{2}$) represents the space of
Hardy
class functions\footnote{
${\cal H}_{+}^{p}$ consists of all complex analytic functions $G_{+}(E+iy)$
defined on the upper half of the complex plane which are $p$-integrable for
any fixed $y>0$ and for which
$\mbox{sup}_{y>0} \intreal \left|{G(E+iy)}\right|^{p} < \infty$.
Their boundary values on the real line exist for almost all $E$ and is
$p$-integrable.
${\cal H}_{+}^{p}$ is analogously defined on the lower half plane.
}
from above (respectively below).

Defining $\Delta_{\pm}:=\hil_{\pm}^{2} \cap {\cal S}$, where ${\cal S}$ is
the Schwartz space, we obtain two Gel'fand triplets
\begin{equation}
\Delta_{\pm} \subset \hil_{\pm}^{2} \subset
   \left(\Delta_{\pm}\right)^{\times}.
\end{equation}
By a suitable (non-unitary) \cite[p.\ 69]{e} map these triplets define a pair
of {\rhs}s
\begin{equation}
\Phi_{\pm} \subset \hil \subset \left(\Phi_{\pm}\right)^{\times}.
\end{equation}
The physical interpretation of these two triplets will be the following:
$\Phi_{-}$ is the space of the preparations, i.~e.\ the physical in-states,
$\Phi_{+}$ is the space of the registrations, i.~e.\ the detected out-states
of the scattering experiment.

It can be shown, that the restriction of the unitary time evolution operator
$U(t)=e^{i\bar{H}t}$ to $\Phi_{\pm}$ is a $\tau_{\Phi}$-continuous operator
with $U(t)\Phi_{+} \subset \Phi_{+}$ only for $t\geq 0$ and analogously a
$\tau_{\Phi}$-continuous operator with
$U(t)\Phi_{-} \subset \Phi_{-}$ only for $t\leq 0$.
This results in a pair of evolution semigroups instead of the ordinary
unitary evolution group $U(t)$ which is the mathematical manifestation of
irreversibility.

\section{\QM\ and the Arrow of Time}
\label{sec:BohrLudwig}

In the conventional description of \qm\ irreversibility is not accounted for
intrinsically and the time evolution is given by the unitary
one-parameter group of operators $U(t)=e^{iHt}$ which is generated by the
Hamiltonian $H$ and well-defined for $-\infty < t <\infty$.
Instead, an arrow of time is considered to be a result of an irreversible act
of measurement and was postulated by von Neumann by a formalism which is
known as the ``collapse of the wave function'' \cite{k,l}:\\
In usual \qm\ a state $W$ changes according to
\begin{equation}
W(t)=U^{\dag}(t)W(0)U(t)
\end{equation}
where $U(t)=e^{iHt}$ is a unitary one-parameter group of operators generated
by the Hamiltonian $H$.
As the time evolution is unitary, there is no arrow of time and we have
\begin{equation}
S[W(t)]\equiv -k \mbox{TR}\left[W(t)ln\left( W\left(t\right)\right)\right]
= -k \mbox{Tr}\left[W(0)ln\left( W\left(0\right)\right)\right]
= S[W(0)].
\end{equation}
Now, according to the postulate of the collapse of the wave function, the
state changes as a result of an ``idealized measurement of the first kind''
such that the state $W$  in which the observable $B=\sum_{i}b_{i}E_{b_{i}}$
is measured collapses into $W'$ according to
\begin{equation}
W\stackrel{\mbox{collapse}}{\longrightarrow} W'
= \sum_{b_{i}} E_{b_{i}}W E_{b_{i}}
 \stackrel{\mbox{reading of result $b_{f}$ }}{\longrightarrow}
E_{b_{f}}W E_{b_{f}}.
\end{equation}
Only $S[W']\geq S[W]$ can be shown, and the measurement is considered to be
the cause of
irreversibility.
It is important to notice that the ``collapse of the wave function'' is a
mathematical idealization expressing the fact that measurement affects the
states and that an immediate repetition of a measurement should lead to the
same result.

The arrow of time we will present in the following is of a different nature
and is located on the microphysical level.
It manifests itself as exponential decay and can be formulated in terms of
states and observables.
The same idea has already been  used by Ludwig \cite {i} who however
formulated this arrow of time in terms of an experimental apparatus only and
extrapolated the time evolution into the past when he transcribed the
observational facts into the mathematical theory.
This was necessary because his Hilbert space formulation of \qm\ does not
allow for an intrinsic \qmal\ arrow of time according to the Bohr-Ludwig
interpretation of \qm.
The division of an experiment into a preparation apparatus, which prepares
physical states for the experiment, and a registration apparatus, which
measures properties of the microsystem, is crucial.
The microsystem itself is the agent by which the preparation apparatus acts
on the registration apparatus.
Whereas the state was prepared by the preparation apparatus the measurements
of the observables $F$ is done by the registration apparatus and the
expectation value represents the average value of an ensemble of
measurements:
\begin{equation}
\mbox{Tr}(WF) = \sum_{\phi} (\phi,F\phi)
      = \sum_{\phi} <\phi | \psi > <\psi | \phi>.
\end{equation}
Here, we choose the special case that the statistical operator $W$ is given
by the vector $\phi$ describing a pure state and the self-adjoint operator
$F$ is the projection operator $|\psi><\psi |$ describing the special
observable called decision observable or property.

Whereas the apparatuses are not described by \qm, the measured value obtained
by the measuring apparatus is predicted by quantum theory.
We observe the dynamics of the microsystem as the time translation of the
registration procedure relative to the preparation procedure.
Thus, if a time direction is distinguished for the time translation of the
registration apparatus relative to the preparation apparatus then this is a
distinguished direction for the time evolution of the microphysical system
described by \qm.
In terms of the apparatus it is possible to describe the arrow of time
independently of a mathematical theory.
We call this arrow the preparation $\rightarrow$ registration arrow and
characterize it by the following statement \cite{i}:
\begin{quote}
Time translation of the registration apparatus relative to the preparation
apparatus makes sense only by an amount $\tau\geq 0$.
\end{quote}
Translated into the \qmal\ notions of the mathematical theory, this arrow is
formulated as:
\begin{quote}
An observable $| \psi(\tau ) >< \psi(\tau) | $ --- in particular a projection
operator --- can be measured in a state $\phi = \phi (0)$ only after the
state has been prepared, i.~e.\ for $\tau \geq 0$.
\end{quote}
This is the formulation in the Heisenberg picture, in the Schr{\"o}dinger
picture it is given as:
\begin{quote}
The state $\phi (t)$ must be prepared before an observable
$|\psi ><\psi | = | \psi(0) >< \psi(0) |$ can be measured in that state,
i.\ e.\ at $t\leq 0$.
\end{quote}
The latter two formulations of the state $\rightarrow$ observable arrow of
time concern the theoretical description of our arrow of time; we call it the
\qmal\ arrow of time.
These formulations should be equivalent to the first version which rather
concerns the experimental implications of the \qmal\ arrow of time.
All version are general expressions of causality and none of them has
anything to do with the change of the state due to measurement, i.~e.\ the
``collapse of the wave function''.

In the next section we shall apply this \qmal\ arrow of time to a resonance
scattering process.
Before we can give a mathematical formulation for this \qmal\ arrow of time
we will have to introduce the standard formalism of scattering theory and
refine it to the Rigged Hilbert Space formulation of \qm.

\section{Mathematical Formulation of the Quantum Mechanical Arrow of Time
in a Resonance Scattering Experiment}

\subsection{Resonances of the $S$-Matrix and Gamow Vectors}
\label{sec:mathform}

The design of a scattering experiment is the following:
A mixture of initial states $\phi^{in}$ is prepared and evolves according to
the free Hamiltonian $K$
\begin{equation}
\phi^{in}(t)=e^{-iKt/\hbar}\phi^{in}.
\end{equation}
The evolution through the interaction region is given by the Hamiltonian
$H=V+K$.
The result is a state $\phi^{out}$,
\begin{equation}
\phi^{out}(t)=S \phi^{in}(t), \hspace{1em} S=\Omega^{-\dagger}\Omega^{+}
\end{equation}
where
\begin{eqnarray}
\Omega^{+}\phi^{in}(t)\equiv \phi^{+}(t)=e^{-iHt/\hbar}\phi^{+} =
 \Omega^{-}\phi^{out}(t) \\[.25cm]
\Omega^{-}\psi^{out}(t)\equiv \psi^{-}(t) = e^{-iHt/\hbar}\psi^{-},
\end{eqnarray}
and $\Omega^{+}$ and $\Omega^{-}$ are the M{\o}ller wave operators.
The state vector $\phi^{in}$ is determined by the preparation apparatus.
The state vector $\phi^{out}$ is determined by the preparation apparatus
{\em and} the dynamics.
The detector does not register $\phi^{out}$, but a property
$|\psi^{out}><\psi^{out}|$, where $\psi^{out}$ is determined by the
registration apparatus.
The probability to measure the observable $|\psi^{out}><\psi^{out}|$ in the
state $\phi^{in}$ is then given by
$\left| <\phi^{in}|\psi^{out}> \right|^{2}$ which is the modulus square of
the $S$-matrix.

The entries of the $S$-matrix are given by the elements
\begin{equation}
 \left( \psi^{out}(t) , \phi^{out}(t) \right) =
 \left( \psi^{out}(t) , S\phi^{in}(t) \right),
\end{equation}
which, for {\em all} $t$, can be given by
\begin{equation}
\label{eq:s-matrix}
\begin{array}{rl}
 \left( \psi^{out}(t), S\phi^{in}(t) \right)
 &= \left( \Omega^{-} \psi^{out}(t), \Omega^{+} \phi^{in}(t) \right)\\[.25cm]
 & = \left( \psi^{-}(t), \phi^{+}(t) \right) \\[.25cm]
 & = \int_{\sigma (H)} dE < \psi^{-} | E^{-}> S_{I}(E+i0) <{}^{+}E |
\phi^{+}>.
\end{array}
\end{equation}
\begin{figure}
\begin{center}
\unitlength=1.00mm
\special{em:linewidth 0.4pt}
\linethickness{0.4pt}
\begin{picture}(149.00,125.00)
\put(10.00,125.00){\line(0,-1){90}}
\linethickness{.1pt}
\multiput(10,38)(0,4){22}{\line(-1,-1){3.6}}
\linethickness{.4pt}
\put(5.00,80.00){\line(1,0){65}}
\bezier{52}(10.00,80.00)(19.00,83.00)(21.00,86.00)
\bezier{64}(21.00,86.00)(26.00,91.00)(29.00,99.00)
\bezier{64}(29.00,99.00)(32.00,107.00)(37.00,107.00)
\bezier{64}(37.00,107.00)(42.00,107.00)(45.00,99.00)
\bezier{64}(45.00,99.00)(48.00,91.00)(53.00,86.00)
\bezier{52}(53.00,86.00)(55.00,83.00)(64.00,80.00)
\put(26.00,92.00){\vector(1,2){2.00}}
\put(38.00,91.00){\circle{8.00}}
\put(38.00,91.00){\circle*{0.30}}
\put(34.00,92.00){\vector(0,-1){2.00}}
\put(12.00,80.00){\oval(4.00,4.00)[lb]}
\put(12.00,78.00){\line(1,0){57}}
\put(32.00,72.00){\makebox(0,0)[cc]{$\underline{\sigma(\bar{H})}$}}
\put(30.00,55.00){\makebox(0,0)[cc]{E (first sheet)}}
\put(30.00,110.00){\makebox(0,0)[cc]{E (second sheet)}}
\put(38.00,91.00){\line(1,0){0}}
\put(39.00,83.00){\makebox(0,0)[cc]{$z_{R}^{*}$}}
\put(20.00,92.00){\makebox(0,0)[cc]{$C_{+}$}}
\put(89.00,125.00){\line(0,-1){90}}
\linethickness{.1pt}
\multiput(89,38)(0,4){22}{\line(-1,-1){3.6}}
\linethickness{.4pt}
\put(84.00,80.00){\line(1,0){65}}
\bezier{52}(89.00,80.00)(98.00,77.00)(100.00,74.00)
\bezier{64}(100.00,74.00)(105.00,69.00)(108.00,61.00)
\bezier{64}(108.00,61.00)(111.00,53.00)(116.00,53.00)
\bezier{64}(116.00,53.00)(121.00,53.00)(124.00,61.00)
\bezier{64}(124.00,61.00)(127.00,69.00)(132.00,74.00)
\bezier{52}(132.00,74.00)(134.00,77.00)(143.00,80.00)
\put(105.00,68.00){\vector(1,-2){2.00}}
\put(117.00,69.00){\circle{8.00}}
\put(117.00,69.00){\circle*{0.30}}
\put(113.00,68.00){\vector(0,1){2.00}}
\put(91.00,81.00){\oval(4.00,4.00)[lt]}
\put(91.00,83.00){\line(1,0){55}}
\put(109.00,40.00){\makebox(0,0)[cc]{E (second sheet)}}
\put(109.00,110.00){\makebox(0,0)[cc]{E (first sheet)}}
\put(118.00,63.00){\makebox(0,0)[cc]{$z_{R}$}}
\put(111.00,88.00){\makebox(0,0)[cc]{$\overline{\sigma(\bar{H})}$}}
\put(99.00,68.00){\makebox(0,0)[cc]{$C_{-}$}}
\end{picture}
\caption{Paths of integration\label{fig:intpath}}
\end{center}
\end{figure}
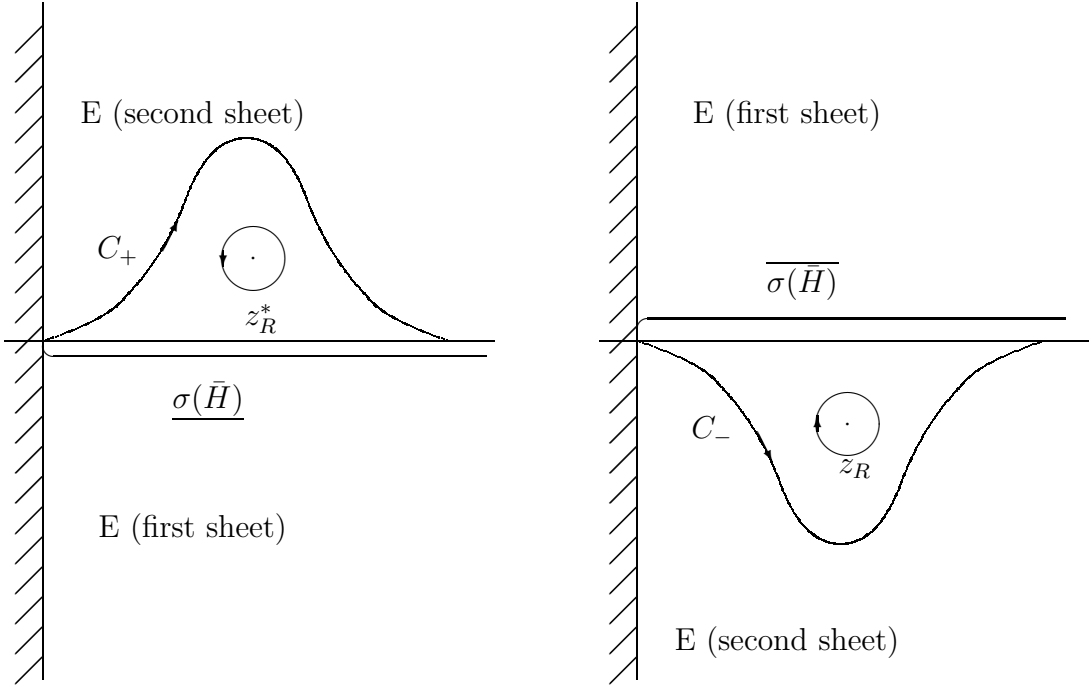
Here $S_{I}(E+i0)$ indicates integration along the upper rim of the cut along
the real axis in the first sheet.
Since $S_{I}(E+i0) = S_{II}(E-i0)$ this integration  along that sheet can be
realized by the integration over the lower rim of the cut in the second
sheet.
The $+$ preceding $E$ in $<{}^{+}E|$ and the $-$ following $E$ in $|E^{-}>$
underline this fact.

In order to define now the Gamow vector in terms of the $S$-matrix we
consider the simplest model and we make the following assumptions (if there
are more resonance poles and also bound states the arguments are easily
generalized):
\begin{itemize}
\item There are no bound or virtual states.
\item The spectrum $\sigma (H)$ of $H=K+V$ is given by the positive real axis
which implies that there is a cut from $0$ to $\infty$ on the Riemann surface
for the analytic continuation $S(z)$ of the $S(E)$-matrix .
\item There is a pair of poles on the second sheet of the complex energy
surface of the $S$-matrix $S(z)$ at the pole position:
\begin{equation}
z_{R}=E_{R} -i\frac{\Gamma}{2} \hspace{.5em},
\hspace{2em}
z_{R}^{*} = E_{R}+i \frac{\Gamma}{2} \hspace{.5em}.
\end{equation}
\end{itemize}
Physically, the poles in the second sheet of the $S$-matrix are associated
with resonances.
We will associate to these pole positions autonomous resonance states and we
call the corresponding generalized state vectors the Gamow vectors.
To accomplish this we deform the contour of integration in
(\ref{eq:s-matrix}) into the second sheet and obtain
\begin{equation}
\label{eq:deformint}
\label{eq:transint}
\begin{array}{rl}
(\psi^{-}, \phi^{+}) =
& \int_{C_{-}} dz <\psi^{-}|\omega^{-}>
         S_{II}(z) <{}^{+}z|\phi^{+}> \\[.25cm]
& + \oint dz <\psi^{-}| z^{-}> \frac{s_{-1}}{z -z_{R}}
    <{}^{+}z | \phi^{+}>
\end{array}
\end{equation}
where $s_{-1} = i\Gamma$ is the residuum of $S_{II}(z)$ at $z_{R}$ and
$C_{-}$ is the curve in the lower second sheet which is shown in figure
\ref{fig:intpath}.
We also have assumed here, that $< \psi^{-} | z^{-}> = $
$<{}^{-}z|\psi^{-}>^{*}$
and $<{}^{+}z|\phi^{+}>$ are analytic functions in the lower half plane of
the second sheet of the Riemann surface.
As the first integral is not related to the resonance we will ignore this
term in the following.

The second integral in (\ref{eq:deformint}) can now be evaluated by means of
the Cauchy Theorem, which yields
\begin{equation}
\label{eq:intpath}
<\psi^{-} | z_{R}^{+}> \left( 2\pi\Gamma\right)
<{}^{+}z_{R} | \phi^{+}> =
\oint dz <\psi^{-}| z^{-}><{}^{+}z | \phi^{+}>
      \frac{i\Gamma}{z -z_{R}}
  \hspace{1em}.
\end{equation}
We now demand further, that the second integral of (\ref{eq:deformint}) is
given by a Breit-Wigner integral, i.~e.\ that (\ref{eq:intpath}) takes the
form
\begin{equation}
\label{eq:titchg-}
\begin{array}{l}
<\psi^{-}|z_{R}^{-}> < {}^{+}z_{R}| \phi^{+}>2\pi\Gamma = \\[.25cm]
\hspace{5em}\int_{-\infty_{II}}^{+\infty} dE
<\psi^{-}|E^{-}><{}^{+}E|\phi^{+}>
  \frac{i\Gamma}{E-(E_{R} -i\frac{\Gamma}{2})}
\hspace{1em}.
\end{array}
\end{equation}
The reason for this requirement is that we want to associate the resonance
``state'' vector with the pole of the $S$-matrix and the resonance state
vector should have a Breit-Wigner energy distribution.

In order to deform the contour of integration from the one in
(\ref{eq:intpath}) into  an integration from $-\infty$ to $+\infty$ as in
(\ref{eq:titchg-}), we need to make the following requirement of the energy
wave functions:
\begin{eqnarray}
\label{eq:hardy+}
<E|\psi^{out}> = <{}^{-}E|\psi^{-}> \in {\cal S} \cap {\cal H}_{+}^{2}
\hspace{1em} \mbox{or } <\psi^{-}|E^{-}> \in {\cal S}\cap{\cal H}_{-}^{2}
\\[.25cm]
\label{eq:hardy-}
<E|\phi^{in}> = <{}^{+}E|\phi^{+}> \in {\cal S} \cap {\cal H}_{-}^{2}
\hspace{1em} \mbox{or } <\phi^{+}|E^{+}> \in {\cal S}\cap{\cal H}_{+}^{2}
\end{eqnarray}
because then
$<\psi^{-}|E^{-}><{}^{+}E|\phi^{+}> \in {\cal S} \cap \hil_{-}^{2}$
and (\ref{eq:titchg-}) is a special case of the Titchmarsh theorem (cf.\
Appendix).
In here ${\cal H}_{+}^{2}$ means Hardy class from above and ${\cal
H}_{-}^{2}$ means Hardy class from below.

If the conditions (\ref{eq:hardy+}) and (\ref{eq:hardy-}) are fulfilled, we
say that $\psi^{-}$ is a ``very well-behaved vector from above'' and
$\phi^{+}$ a ``very well-behaved vector from below'' and write this as
\begin{equation}
\psi^{-}\in \Phi_{+}, \hspace{1em} \phi^{+}\in \Phi_{-}.
\end{equation}
This means the space of very well behaved vectors from below, $\Phi_{-}$,
describes the state vectors prepared by the preparation apparatus, and the
space of very well behaved vectors from above, $\Phi_{+}$, describes the
observables detected by the registration apparatus.
We know that a more careful investigation would show that
$\Phi_{+} \cap \Phi_{-} \neq \{ 0 \}$ (zero vector) \cite{e}.
We shall see below, that conditions (\ref{eq:hardy+}) and (\ref{eq:hardy-})
are exactly the mathematical statement describing the \qmal\ arrow of
time.
It will thus come as no surprise that the time development of the
Gamow vectors is given by two semigroups instead of a unitary evolution
group.

It may seems curious that we label the vectors $\psi^{-}$, $\phi^{+}$ by
superscripts that are opposite to the subscripts by which we label their
spaces $\Phi_{+}$, $\Phi_{-}$.
The reason for this is that the labelling of the vectors comes from the
convention of scattering theory developed by physicists and the labelling of
the spaces comes from the convention for the Hardy class spaces
$\hil_{+}^{2}$ and $\hil_{-}^{2}$ used by mathematicians.
Both subjects were developed independently and without knowing of each other.
The agreement between $\{ \psi^{-} \}$ and $\Phi_{+}$ as well as
$\{\phi^{+} \}$ and $\Phi_{-}$ is an example of ``the unreasonable
effectiveness of mathematics in the natural sciences'' (E. P. Wigner).

We are now able to define the Gamow vectors $|{z_{R}}^{-}>$ and
$|{z_{R}^{*}}^{+}>$ associated with the poles at $z_{R}$ and $z_{R}^{*}$.
In analogy to (\ref{eq:titchg-}) we apply the Titchmarsh Theorem to the
function
$<\psi^{-}|E^{-}>\in {\cal H}_{-}^{2}$ with value $<\psi^{-}| {z_{R}}^{-}>$
at $z_{R}$ to obtain:
\begin{equation}
\label{eq:gamow-}
\begin{array}{r}
<\psi^{-}|{z_{R}}^{-}> =
 -\frac{1}{2\pi i} \int_{-\infty_{II}}^{+\infty} dE
   <\psi^{-}|E^{-}>  \frac{1}{E-z_{R}} \\[.25cm]
   \mbox{for all } <{}^{-}E|\psi^{-}>\in {\cal S} \cap {\cal H}_{+}^{2}
\end{array}
\end{equation}
and to the function $<\phi^{+}|E^{+}>\in {\cal H}_{+}^{2}$ with value
$<\phi^{+}| {z_{R}^{*}}^{+}>$ at $z_{R}^{*}$ to get:
\begin{equation}
\label{eq:gamow+}
\begin{array}{r}
<\phi^{+}|{z_{R}^{*}}^{+}> =
   \frac{1}{2\pi i} \int_{-\infty_{II}}^{+\infty} dE
   <\phi^{+}|E^{+}>  \frac{1}{E-z_{R}^{*}} \\[.25cm]
   \mbox{for all } <{}^{+}E|\phi^{+}>\in {\cal S} \cap {\cal H}_{-}^{2}.
\end{array}
\end{equation}
Notice, that in equation(\ref{eq:gamow-}), integration is taken along the
lower rim on the second sheet whereas in equation (\ref{eq:gamow+}) along the
upper rim on the second sheet.

It is clear that $|{z_{R}}^{-}>$ defined by (\ref{eq:gamow-}) is an element
of $\Phi_{+}^{\times}$ and $| \zRsp> $ defined by (\ref{eq:gamow+}) is
an element of $\Phi_{-}^{\times}$.
These generalized vectors (functionals) are the two kinds of Gamow vectors.
It is crucial that the integration in (\ref{eq:gamow-}) and (\ref{eq:gamow+})
extends from $-\infty$ to $+\infty$ though the spectrum of $H$ is bounded
from below.

Replacing $\psi^{-}$ by $H\psi^{-}$ (respectively $\phi^{+}$ by $H\phi^{+}$)
in equation (\ref{eq:gamow-}) (respectively equation (\ref{eq:gamow+})) and
using
$<H\psi^{-} | E^{-}> = E <\psi^{-}| E^{-}>$
(respectively $<H\phi^{+} | E^{+}> = E <\phi^{+}| E^{+}>$)
we can infer that the Gamow vector $|{z_{R}}^{-}>$
(respectively $|{z_{R}^{*}}^{+}>$) is a generalized eigenvector of $H$ with
the
eigenvalue $z_{R}$ (respectively $z_{R}^{*}$):
\begin{eqnarray}
<H\psi^{-} | z_{R}^{-} > \equiv < \psi^{-} | H^{\times} | z_{R}^{-} >
 = z_{R} < \psi^{-} | z_{R}^{-} >
\\[.2cm]
\mbox{ or } \;
H^{\times} | z_{R}^{-} > = z_{R} | z_{R}^{-} >;\;
|\zRm > \in \Phi_{+}^{\times} \nonumber\\[.25cm]
<H\phi^{+} | z_{R}^{* -} > \equiv < \phi^{+} | H^{\times} | \zRsp >
 = z_{R}^{*} < \phi^{+} | \zRsp >
\\[.2cm]
\mbox{ or } \;
H^{\times} | \zRsp > = z_{R}^{*} | \zRsp >;\;
|\zRsp > \in \Phi_{-}^{\times} .\nonumber
\end{eqnarray}

\subsection{Time Evolution of the Gamow Vectors}
\label{sec:gamow}
In the usual Hilbert space $\hil$, the time evolution, given by $e^{iHt}$,
is a continuous operator for all $t\in \real$.
The time evolution operators of the Gamow vectors
$|{z_{R}}^{-}> \in \Phi_{+}^{\times}$ and
$|{z_{R}^{*}}^{+}>\in\Phi_{-}^{\times}$
are the conjugate operators of
$U(t)|_{\Phi_{+}}=e^{iHt}|_{\Phi_{+}}$ and
$U(t)|_{\Phi_{-}}=e^{iHt}|_{\Phi_{-}}$, respectively.
The conjugate operators can only be defined if the operators themselves are
continuous operators with respect to the topology in $\Phi$.
In the case of $U(t)|_{\Phi_{+}}$ this is only the case for $t\geq 0$ and we
denote the conjugate by ``${e_{+}}^{-iHt}$''.
On the other hand the operator $U(t)|_{\Phi_{-}}$ is a continuous operator
in $\Phi$ only for $t\leq 0$ and here, the conjugate is denoted by
``${e_{-}}^{-iHt}$''.
Thus, ``$e_{+}^{-iHt}$'' exists only for $t\geq 0$ and ``$e_{-}^{-iHt}$''
exists only for $t\leq 0$.

Now, the time evolution of the Gamow vectors $|{z_{R}}^{-}> \in \Phi_{+}$
can be calculated to be
\begin{equation}
\begin{array}{rl}
\left< e^{iHt}\psi^{-} | {z_{R}}^{-} \right>
\equiv &
\left<{\psi^{-} \left| \mbox{``}{e_{+}}^{-iHt}\mbox{''} \right| {z_{R}}^{-}
}\right>
\\[.25cm]
= & e^{-iE_{R}t} e^{\frac{\Gamma}{2}t} \left<\psi^{-}|{z_{R}}^{-} \right>
\mbox{ for all } \psi^{-}\in \Phi_{+}, \mbox{and for } t\geq 0
\end{array}
\end{equation}
and the time evolution of $|{z_{R}^{*}}^{+}>$  is calculated to be:
\begin{equation}
\begin{array}{rl}
\left< e^{iHt}\phi^{+} | \zRsp \right>
\equiv &
\left<{\phi^{+}
\left| \mbox{``}{e_{-}}^{-iHt}\mbox{''} \right|
{z_{R}^{*}}^{+} }\right>
 \\[.25cm]
= &
e^{-iE_{R}t} e^{\frac{\Gamma}{2}t} \left<\phi^{+}|{z_{R}^{*}}^{+} \right>
\mbox{ for all } \phi^{+}\in \Phi_{-}, \mbox{and for }t\leq 0
\hspace{.5em}.
\end{array}
\end{equation}
Thus the vector $|\zRm>$ is an exponentially decaying state vector.
If $\left| < \psi^{-} | z_{R}^{-} > \right|^{2}$ is the probability of
finding the Gamow state $|z_{R}^{-}>$ with the detector that registers
the property $|\psi^{-}><\psi^{-}|$ at $t=0$ then
\begin{equation}
\left| < \psi^{-} | e^{-iHt} | \zRm > \right|^{2} =
e^{-\Gamma t} \left| < \psi^{-} | \zRm > \right|^{2}
\end{equation}
is the probability to find this Gamow state at the time $t$.
Thus the probability decreases exponentially towards $0$ as $t$ grows.
Analogously, the increasing state $|{z_{R}^{*}}^{+}>$ is defined for
$t\leq 0$ with probability increasing exponentially from $0$ in the distant
past to $\left| < \phi^{+} | \zRsp > \right|^{2} $ at $t=0$.

This behaviour can be summarized by saying that $|{z_{R}}^{-}>$ and
$|{z_{R}^{*}}^{+}>$ have an intrinsic arrow of time which evolves only from
$0$ to $\infty$ or from $-\infty$ to $0$, i.~e.\ $|{z_{R}}^{-}>$ does not
grow and $|{z_{R}^{*}}^{+}>$ does not decay.
\smallskip

Thus a microphysical system, described by such a state vector, is equipped
with an arrow of time.

\subsection{A New Generalized Basis Vector Expansion}

The basis vector expansions, of which the Nuclear Spectral Theorem
(\ref{eq:nst}) is an example, provide very important tools for \qm.
They are generalizations of the well known basis vector expansion
$\vec{X}=\sum_{i=1}^{3} x^{i} \vec{e}_{i}$ in the three dimensional Euclidean
space $\real^{3}$.
In the $N$-dimensional complex space this expansion in terms of eigenvectors
$e_{i}$, $i=1,\ldots,N$ of any self-adjoint linear operator $H$ is called the
fundamental theorem of the linear algebra.
In the infinite dimensional Hilbert space $\hil$ this expansion
\begin{equation}
\label{eq:eigvecexp}
\hil \owns h = \sum_{n=1}^{\infty} \sum_{b} |E_{n},b) ( b,E_{n}|h);\;
|E_{n},b)=e_{i} \in \hil
\end{equation}
is only correct for a subset of self-adjoint operators $H$ (compact).
Here we have called $b$ the degeneracy index:
$H|E_{n_{j}},b) = E_{n} | E_{n_{j}},b)$.
For any arbitrary self-adjoint operator a generalization of
(\ref{eq:eigvecexp}) holds, which is given by the fundamental Nuclear
Spectral Theorem of the generalized Dirac basis vector expansion
(\ref{eq:nst}).
(To take degeneracy into account the label $b$ should be added: $|E,b>$
and a sum should be taken over all values of the quantum numbers $b$.)
The set of generalized eigenvalues $\{E_{n}, E\}$  is called the spectrum
(we have assumed that the spectrum of $H$ was absolutely continuous and
discrete).

With the help of the Gamow vectors we can generalize the Dirac basis vector
expansion further.
For a self-adjoint Hamiltonian $H=K+V$ for which there are only $N$ simple
poles of the $S$-matrix at $z_{R_{i}}=E_{i}-i\frac{\Gamma_{i}}{2}$,
$i=1,\ldots,N$ one obtains (from a generalization of equation
(\ref{eq:deformint}) from 1 to $N$ poles) a new generalized basis vector
expansion
\begin{equation}
\label{eq:nbve}
\begin{array}{rcl}
\phi^{+} & = & \sum_{i=1}^{N} \sum_{b}
|z_{R_{i}}^{-}, b > (2 \pi \Gamma_{i}) <{}^{+}b,z_{R_{i}} | \phi^{+} >
\\[.25cm]
& & +
\int_{0}^{-\infty} \sum_{b} dE |E,b{\,}^{+}> <{\,}^{+}b,E|\phi^{+}>
\mbox{ for every } \phi^{+} \in \Phi_{-}.
\end{array}
\end{equation}
We have ignored the sum over the discrete spectrum
$\sum_{n=1}^{\infty} |E_{n})(E_{n}|\phi^{+}) $,
corresponding to bound states.

In contrast to Dirac's basis vector expansion (\ref{eq:nst}) which holds for
every $\phi \in \Phi$, the new basis vector expansion (\ref{eq:nbve}) holds
only for
$\phi^{+} \in \Phi_{-} \subset \Phi = \Phi_{+} + \Phi_{-}$
and contains generalized eigenvectors
$|{z_{R_{i}}}^{-}> \in \Phi_{+}^{\times}$.
It means that a vector $\phi^{+}$ representing a state (ensemble) prepared
by
a preparation apparatus of a scattering experiment can be expanded into a
superposition of Gamow vectors
representing exponentially decaying resonance ``states'' {\em and} a
background term
\begin{equation}
\phi_{bg}^{+} = \int_{0}^{-\infty} \sum_{b} dE
|E,b{\;}^{+}> <{\;}^{+}b,E | \phi^{+}>; \;
<{}^{+} b,E | \phi^{+}> \in \hil_{-}^{2} \cap {\cal S}.
\end{equation}
The integration in the background term is taken in the second sheet
along the negative real axis
(and the values of $< {}^{+}E | \phi^{+}>$ can be calculated from the
physical values $<{}^{+}E | \phi^{+}> = <E | \phi^{in}>$ with
$0 \leq E < +\infty $ given by the energy distribution of the incoming beam
using the von Winter Theorem \cite{e}) and could have been replaced by
integration along other equivalent contours.
The result (\ref{eq:nbve}) shows that resonance ``states'' can take a very
similar position as bound states in the basis vector expansion of states
$\phi^{+} \in \Phi_{-}$ prepared by a scattering experiment.
These state vectors $\phi^{+}$ can indeed be given by a superposition of
exponentially decaying resonance states
$\psi_{i}^{D} = |{z_{R_{i}}}^{-}> \sqrt{2\pi\Gamma_{i} }$ (except for a
background term which one would try to make as small as possible), as used
so often in phenomenological discussions of physical problems.

\subsection{The Quantum Mechanical Arrow of Time and Resonance Scattering}

We now want to discuss the connection between the Gamow vector's intrinsic
arrow of time and the more general \qmal\ arrow of time that we had
introduced in section \ref{sec:BohrLudwig} and which was nothing else but a
general expression of causality.
For this purpose, we apply that general arrow of section \ref{sec:BohrLudwig}
to the resonance scattering experiment.

We choose as $t=0$ the point in time at which the preparation of the state
$\phi^{in}(t)$ is completed and after which the registration of
$|\psi^{out}> <\psi^{out}|$ begins.
Since no preparations take place for $t>0$, the energy distribution given by
the preparation apparatus must vanish and we require therefore that
$<E | \phi^{in}(t)>=0$ for $t>0$ and for all physical values of $E$.
And since as a general result of scattering theory
\begin{equation}
< E | \phi^{in} > = < {}^{+} E | \phi^{+} >
\mbox{ and }
< E | \psi^{out} > = < {}^{-} E | \psi^{-} >,
\end{equation}
 where $|E>$ and $|E^{\pm}>$ are defined by the Lippman-Schwinger equation
\begin{equation}
|E^{\pm} > = |E> + \frac{1}{E-H\pm i0}V |E> = \Omega^{+}|E>
\end{equation}
and by $K|E>=E|E>$ and $H|E^{\pm}>=E|E^{\pm}>$, we formulate this requirement
of the \qmal\ arrow of time as:
\begin{equation}
\label{eq:fourier1}
\begin{array}{rl}
 0 = & \intreal dE <^{+}E | \phi^{+}(t)>
  = \intreal dE <^{+}E | e^{-iHt/\hbar}|\phi^{+}> \\[.25cm]
 = & \intreal dE <^{+}E |\phi^{+}>e^{-iEt/\hbar}
 \equiv {\cal F}(t) \hspace{1em}\mbox{for }t>0.
\end{array}
\end{equation}
Notice that ${\cal F}$ is the Fourier transform of the energy wave function
$<^{+}E | \phi^{+}> =$ \mbox{$ <E|\phi^{in}>$}.

Similarly, since all registrations take place after $t=0$,
\begin{equation}
\label{eq:fourier2}
\begin{array}{rl}
0= & \intreal dE < E | \psi^{out}(t)> \\[.25cm]
 = &  \intreal dE <{ }^{-} E | \psi^{-}>e^{-iEt/\hbar}
 \equiv {\cal G}(t) \hspace{1em}\mbox{for }t<0,
\end{array}
\end{equation}
where ${\cal G}(t)$ is the Fourier transform of $<^{-}E|\psi^{-}>$ (here, $-$
in $<{}^{-}E|$ refers to integration along the lower rim of the first sheet).

The equations (\ref{eq:fourier1}) and (\ref{eq:fourier2}) are the
mathematical formulations of the general \qmal\ arrow of time, namely of ``no
preparations for $t>0$'' and of ``no registrations for $t< 0$'',
respectively.
{}From the mathematical statements (\ref{eq:fourier1}) and (\ref{eq:fourier2})
one can derive the mathematical properties of the spaces $\Phi_{-}$ and
$\Phi_{+}$ in the following way:
As $<{}^{+}E |\phi^{+}>$ as well as $<{}^{-}E|\psi^{-}>$ are the energy
distributions of the experimental apparatuses, they are supposed to be
smooth, well-behaved functions of $E$.
Thus, the  Theorem of Paley-Wiener is applicable, which says that a square
integrable function $G_{+}(E)$ (respectively $G_{-}(E)$) belongs to
${\cal H}_{+}^{2}$ (respectively ${\cal H}_{-}^{2}$) if and only if it is the
Fourier transform of a square integrable function which vanishes on the
interval $(0,\infty)$ (respectively $(-\infty , 0)$).
In our situation this yields
\begin{eqnarray}
<{}^{-}E |\psi^{-}> \in {\cal H}_{+}^{2} \cap {\cal S}
& \mbox{(or } \psi^{-}\in\Phi_{+} \mbox{) } \\[.25cm]
<{}^{+} E |\phi^{+}> \in {\cal H}_{-}^{2} \cap {\cal S}
& \mbox{(or } \phi^{+}\in\Phi_{-} \mbox{). }
\end{eqnarray}
These are precisely the conditions (\ref{eq:hardy+}) and (\ref{eq:hardy-})
which we obtained above from the requirement that the pole term of the $S$-
matrix has a Breit-Wigner energy distribution.
This means that we could have taken the \qmal\ arrow of time in its
mathematical formulation (\ref{eq:fourier1}) and (\ref{eq:fourier2}) as the
starting point and derived from it the Rigged Hilbert Spaces
$\Phi_{+} \subset \hil \subset \Phi_{+}^{\times}$ of observables (or
out-``states'' $\psi^{-}$) and
$\Phi_{-} \subset \hil \subset \Phi_{-}^{\times}$ of states (or
in-states $\phi^{+}$) and their Gamow vectors $|\zRm>$ and $\zRsp>$
respectively.

\section{Exact Golden Rule}

Decaying states can be considered as resonances for which the production
process is ignored.
As an example one can think of the radiative transitions of excited atoms
$A^{\ast}$ to lower states $A$ by
$A^{\ast} \rightarrow A + \gamma$.
Whereas stationary states are rare in physics, decaying states are
numerous and thus constitute an important part of physics.

If the decaying system is isolated, its state $W(t)$ evolves in time
according to the exact Hamiltonian $H$ and is given by
\begin{equation}
W(t)=e^{-iHt}W(0) e^{iHt}, \hspace{1em} \mbox{with }H=K+V \geq 0.
\end{equation}
As registration apparatus we choose detectors which surround $A^{\ast}$.
We suppose that the decay products are far enough from each other so that
they do not interact after the decay.
Then the observables are projection operators $\Lambda$ (or positive
operators) on the space of physical states of the decay products and are
given by
\begin{equation}
 \Lambda = \intrealp dE \sum_{b} | E,b><E,b|
\end{equation}
where $|E,b>$ are eigenvectors of $K$ with eigenvalue $E$ and not
eigenvectors of $H$.

Denote by ${\cal P}(t)$ the transition probability, i.~e.\ the probability
to find the decay product $\Lambda$ in the state $W(t)$.
Then ${\cal P}(t)$ is given by
\begin{equation}
{\cal P}(t) = \mbox{Tr}(\Lambda W(t)).
\end{equation}
Inserting for $W(t)=|\psi_{t}><\psi_{t}|$ with $\psi_{t} \in {\cal H}$ one
can show that ${\cal P}(t)$ is identically zero for all $t$ if it was so for
some time interval in the past \cite{hegerfeldt}.
For the derivative of ${\cal P}(t)$ at $t=0$, i.~e.\ for the initial decay
rate $\dot{\cal P}(t=0)$ however, the standard treatment \cite{goldberg}
leads to an approximate formula given by the Golden Rule, which shows that
${\cal P}(t=0)\neq 0$ as it should be.
Since nothing more than the most fundamental assumptions of Hilbert space
\qm\ enter in this derivation we have to conclude that the transition
probability cannot be derived in the framework of the Hilbert space
formulation in a consistent way.

In contrast to this, in the {\rhs}-formulation one can derive an exact Golden
Rule for ${\cal P}(t)$ and obtain $\dot{\cal P}(t)$ as a derivative of it if
one chooses for the decaying state $W(t)$ the Gamow state given by the vector
$\psi^{G} = | \zRm > f $ (with $f$ being a normalization constant):
\begin{equation}
W(t)=|\psi^{G}(t) >< \psi^{G} (t)| =
e^{-\Gamma t} | \psi^{G}><\psi^{G}|
\mbox { for } t\geq 0 \mbox{ only.}
\end{equation}
Then one obtains
\begin{equation}
\label{eq:prob}
{\cal P}(t) = 1 - e^{-\Gamma t}
  \intrealp dE \sum_{b \neq b_{G}}
     \frac{|<E,b | V| \psi^{G}>|^{2}}
     {(E-E_{R})^{2}+ \left(\frac{\Gamma}{2}\right)^{2}},
\mbox{ for } t\geq 0,
\end{equation}
and from this by differentiation:
\begin{equation}
\dot{\cal P} (t) = e^{-\Gamma t} 2\pi \intrealp dE
   \sum_{b\neq b_{G}}
   \frac{\Gamma}{2\pi}\;
   \frac{\left| < E,b | V | \psi^{G} > \right| ^{2}}
   {\left(E-E_{R}\right)^{2} + \left(\frac{\Gamma}{2}\right)^{2} }
\mbox{ for }t\geq 0
\; .
\end{equation}
{}From this one finds --- using the fact that the probability to find the
decay product at time $t\leq 0$ must be $0$, ${\cal P}(0)=0$ --- that
${\cal P}(t)= 1- e^{-\Gamma t}$ and $\dot{\cal P}(0)=\Gamma$ (i.~e.\ the
initial decay rate is equal to the imaginary part of the resonance pole
position of the $S$-matrix, which had already been shown above in section
\ref{sec:mathform} to be equal to the width of the Breit-Wigner energy
distribution).

In the Born approximation, which is defined by
\begin{equation}
\psi^{G} \rightarrow f^{d},
\hspace{2em}
E_{R} \rightarrow E_{d},
\hspace{2em}
\frac{\Gamma}{2 E_{R}} \rightarrow 0,
\end{equation}
where $K f^{d}=E_{d} f^{d}$ is the eigenvector of the free Hamiltonian
approximating the decaying state $\psi^{G}$, one obtains
\begin{equation}
\dot{\cal P} (0) = \Gamma =
   2\pi \intrealp dE
   \sum_{b\neq b_{G}} \left| < E,b | V | f^{d} > \right| ^{2}
   \delta (E-E_{d})\hspace{1em}.
\end{equation}
This is the standard Golden Rule for the transition from the excited but
non-interacting state $f^{d}$ into the mixture of non-interacting decay
products.

\section{Summary and Conclusions}

Quasistationary microphysical systems and their resonances can be described
by Gamow vectors which are generalized eigenvectors in a suitably chosen
space of self-adjoint Hamiltonians with complex eigenvalues.
The description of Gamow vectors is not possible in the Hilbert space,
but the Rigged Hilbert Space allows to define Gamow vectors in
$\Phi^{\times}$.
These vectors are associated to the resonance poles of the $S$-matrix.

The time evolution of the Gamow vectors is governed by semigroups of
operators in the spaces $\Phi_{\pm}^{\times}$ and thus
displays irreversibility.
This is a microphysical arrow of time which is built in the microphysical
systems described by the Gamow vectors.
The conjugate semigroups of operators in $\Phi_{\pm}$ describe a general
\qmal\ arrow of time which is the theoretical description of the
``preparation $\rightarrow$ registration'' arrow of time of the experimental
apparatuses.

This ``preparation $\rightarrow$ registration'' arrow  of time has been known
for some time, but could not be transcribed faithfully from the experimental
observation into the mathematical theory of \qm\ in Hilbert space.
Finally, with the Gamow vector to describe a decaying state, one can derive
a sensible result for the transition probability into non-interacting decay
products and obtain the standard Golden Rule for the transition rate as the
Born approximation.
The Gamow vector provides the link that was missing from the Hilbert space
formulation to connect the theoretically and experimentally defined
quantities
for the decay phenomena.

\section*{Appendix}
\begin{th}[Titchmarsh]
Let $G_{\pm}(z)\in {\cal H}_{\pm}^{p}$.
Then, for any $z=E_{0}+iy$ with $y\grle 0$, one has:
\begin{equation}
G_{\pm}(z)=\pm \frac{1}{2\pi i}\intrealp \frac{G_{\pm}(E)}{E-z} dE
\hspace{1em} \mbox{Im}z \grle 0
\end{equation}
and
\begin{equation}
\intrealp \frac{G_{\pm}(E)}{E-z^{*}}dE =0.
\end{equation}
\end{th}

In our case the function under consideration is
$G_{-}(E)=<\psi^{-}|E^{-}><{}^{+}E | \phi^{+}>$.

Similarly, we obtain
\begin{equation}
\label{eq:titchg+}
\begin{array}{l}
<\phi^{+}|{z_{R}^{*}}^{+}> < z_{R}^{*}| \psi^{-}> = \\[.25cm]
\frac{1}{2\pi i} \int_{-\infty_{II}}^{+\infty} dE
   <\phi^{+}|E^{+}><{}^{-}E | \psi^{-}>
   \frac{1}{E- \left( E_{R}-i\frac{\Gamma}{2} \right)}
  \hspace{1em},
\end{array}
\end{equation}
for $z_{R}^{*}=E_{R}+i\frac{\Gamma}{2}$ and the function
$G_{+}(E)= <\phi^{+} | E^{+}><{}^{-}E|\psi^{-}>$.
Notice, that here the integration takes place along the upper edge of the
real axis in the second sheet.
\nocite{hegerfeldt}

\end{document}